\documentclass[5p]{elsarticle}

\usepackage[cp1250]{inputenc}

\usepackage{hyperref}

\journal{Journal of Luminescence}

\usepackage{epstopdf}
\begin{document}
\begin{frontmatter}
\title{Epitaxial growth and Photoluminescence Excitation spectroscopy of
CdSe Quantum Dots in (Zn,Cd)Se barrier}


\author[a,b]{J. Piwowar}
\author[a]{W. Pacuski}
\author[a]{T. Smole\'nski}
\author[a]{M. Goryca}
\author[a]{A. Bogucki}
\author[a]{A. Golnik}
\author[a]{M. Nawrocki}
\author[a]{P. Kossacki}
\author[a]{J.~Suffczy\'nski\corref{mycorrespondingauthor}}
\address[a]{University of Warsaw, Faculty of Physics, Pasteura 5 St.,
02-093, Warsaw, Poland;}
\address[b]{University of Warsaw, Department of Chemistry, Biological and
Chemical Research Centre, \.{Z}wirki i Wigury 101, Warsaw, Poland}



\cortext[mycorrespondingauthor]{Corresponding author}
\ead{Jan.Suffczynski@fuw.edu.pl}


\begin{abstract}
Design, epitaxial growth, and resonant spectroscopy of CdSe Quantum Dots
(QDs) embedded in an innovative (Zn,Cd)Se barrier are presented. The
(Zn,Cd)Se barrier enables shifting of QDs energy emission down to 1.87~eV,
that is below the energy of Mn$^{2+}$ ions internal transition (2.1~eV).
This opens a perspective for implementation of epitaxial CdSe QDs doped
with several Mn ions as, e. g., the light sources in high quantum yield
magnetooptical devices. Polarization resolved Photoluminescence Excitation
measurements of individual QDs reveal sharp ($\Gamma <$ 150 $\mu$eV) maxima
and transfer of optical polarization to QD confining charged exciton state
with efficiency attaining 26\%. The QD doping with single Mn$^{2+}$ ions is
achieved.
\end{abstract}

\begin{keyword}
quantum dot, exciton, Auger recombination, Photoluminescence Excitation,
CdSe
\end{keyword}

\end{frontmatter}

\section{Introduction}

Selenium based Quantum Dots (QD) act as temperature robust non-classical
light sources \cite{Sebald:APL2002}, what makes them excellent candidates
for applications in optoelectronic and spintronic devices
\cite{Murayama:PRB2007, Dietl:Spintronics2009}. In particular, appealing
properties and functions have been predicted and demonstrated for CdSe QDs
doped with either a single \cite{Kobak:NatureCom2014, Smolenski:PRB2015} or
a density \cite{Worschech:SSciTech2008, Beaulac:NanoLett2008,
Barrows:JPhysChemLett2015} of magnetic ions, like Mn$^{2+}$. The
development of advanced devices involving Mn-doped CdSe QDs and
nanostructures is impeded, however, by Auger-type, non-radiative carrier
recombination introduced by the magnetic dopants
\cite{Abramishvili:SSC1991, Nawrocki:PRB1995, Chernenko:PRB2005}. In the
case of typical Mn-doped CdSe/ZnSe QDs, the QDs emission energy
($\sim$2.3~eV) exceeds the energy of the Mn$^{2+}$ intraionic $^{6}A_{1}
\rightarrow ^{4}T_{1(2)}$ transition (2.1~eV). As a result, the
non-radiative channel of exciton decay related to the ions is efficient
enough to limit significantly or quench completely the excitonic emission
\cite{Worschech:SSciTech2008, Beaulac:NanoLett2008, Chernenko:PRB2005,
Lee:PRB2005, Chekhovich:PRB2007}. The excitonic emission can be, however,
recovered by shifting the QD levels below the energy of the Mn$^{2+}$ ion
transition. In the case of colloidal CdSe QDs such shifting has been
achieved by increasing the nanocrystal size leading to a reduction of the
exciton confinement energy \cite{Beaulac:NanoLett2008} or by a coupling of
the QDs to a graphene sheet \cite{Kanodarwala:JofLumin2014}. A respective
shifting method has been still missing, however, for epitaxial CdSe QDs.

Here, we design and demonstrate epitaxially grown CdSe QDs with energy of
confined levels lowered with respect to a typical case of CdSe/ZnSe QDs by
lowering of the barrier energy. We implement (Zn,Cd)Se barrier exploiting
the fact that (Zn,Cd)Se band gap decreases with increasing Cd content
\cite{Gupta:ThinSF1995, Mourad:PRB2010}. We show that for Cd content
attaining 30 \%, the CdSe QDs transitions are shifted down to 1.87 eV, that
is far below 2.1~eV. This is beneficial for high quantum yield
magnetooptical devices involving CdSe/(Zn,Cd)Se QDs doped with several
Mn$^{2+}$ ions.

Due to a shortage of tunable and stable excitation sources operating in
590~nm - 620~nm spectral range, Photoluminescence Excitation studies of the
selenium based QDs have been primarily limited to ensembles of
\cite{Beaulac:NanoLett2008, Norris:PRB1996, Kagan:PRL1996} or individual
\cite{Htoon:PRL2004} colloidal type CdSe QDs so far.

We demonstrate an optical addressing of individual CdSe QDs using a
quasi-resonant excitation \cite{Puls:PRB1999, Flissikowski:PRL2001,
Scheibner:PRB2003} with a standard Rhodamine 590 laser. We find an
efficient transfer of the optical polarization for a charged exciton
confined in the QD. We incorporate single Mn$^{2+}$ ions to the QDs by a
low density doping at the sample growth stage. When combined with a
possibility of resonant excitation, it makes the studied QDs promising for
efficient optical orientation of a single magnetic ions spin
\cite{Kobak:NatureCom2014, Smolenski:PRB2015, Goryca:PRL2014}. Moreover,
the emission range of the presented QDs provides a perspective for their
implementation as temperature robust light sources and as building blocks
of lasers operating in the short wavelength transmission window of plastic
optical fibers \cite{Murofushi:POF1996}.

\section{Samples and experiment}

A sample series comprises of twelve structures grown by Molecular Beam
Epitaxy (MBE) on a GaAs (100) substrate followed by a 1.1 $\mu$m thick ZnSe
buffer and 1~$\mu$m thick (Zn,Cd)Se barrier layer. The CdSe QDs are formed
out of 2-3 monolayers of CdSe deposited in atomic layer epitaxy mode,
100~nm below the sample surface. The cadmium content in the barrier layer x$_{Cd}$
varies from 0\% to 30\%, depending on a sample from the series. Some of the
samples are doped with a very low density of Mn$^{2+}$ ions in the QDs
layer.

The samples undergo a characterization by $\mu$-Photolumi\\nescence
($\mu$-PL) at T = 10~K excited non-resonantly at E$_{exc}$ = 3.06~eV.
Micro-Photoluminescence Excitation (PLE) experiment at T = 1.8~K employs a
continuous wave dye (Rhodamine-590) laser tunable between 2.21 eV and 2.06
eV (560~nm-600~nm) as an excitation source. The excitation power is
live-stabilized by a noise eater. The excitation beam is focused by a
microscope objective down to 1~$\mu$m or 0.5~$\mu$m, in the case of
$\mu$-PL characterization and PLE studies, respectively. Magnetic field of
up to 10~T is applied in Faraday configuration ($\vec{k} \parallel
\vec{B}$). The polarization resolved signal is detected using a
Peltier-cooled CCD camera coupled to a 0.5~m grating spectrometer (100
$\mu$eV of overall spectral setup resolution).

\section{Tuning of CdSe/(Zn,Cd)Se Quantum Dots emission energy}
\begin{figure}[!t]
\includegraphics[width=0.9\linewidth]{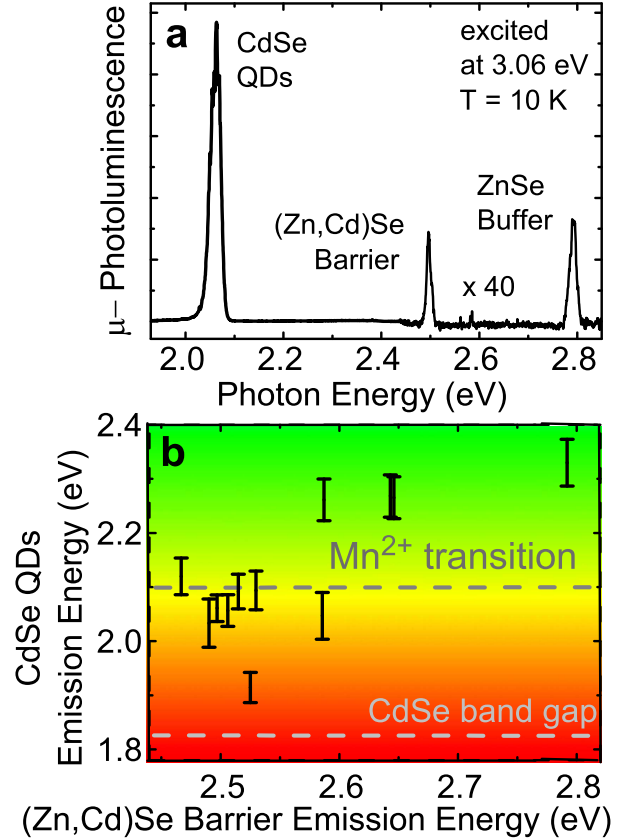}
\caption{(a) Typical $\mu$-Photoluminescence spectrum of the studied sample
excited nonresonantly at 3.06 eV. The emission from (Zn,Cd)Se barrier and
ZnSe buffer shown in $\times$ 40 magnification for a better visibility. (b) Energy
ranges covered by CdSe/(Zn,Cd)Se QDs emission as a function of (Zn,Cd)Se
barrier emission energy determined for the studied sample series. Energies
of bulk CdSe band gap and of Mn$^{2+}$ ions internal transitions are
indicated with horizontal dashed lines for a reference.}
\label{fig:tuning}
\end{figure}

A typical emission pattern of a sample from the studied series is shown in
Figure~\ref{fig:tuning}a). Here, the QDs emission is seen as a band in
range of 2.04-2.08~eV. Much weaker (Zn,Cd)Se barrier and ZnSe buffer
emission is centered at 2.49~eV at 2.79~eV, respectively. In the sample
set, the (Zn,Cd)Se barrier emission energy varies from 2.46~eV to 2.79~eV
(the latter corresponding to the pure ZnSe case), while the central energy
of the QDs ensemble emission band covers the range from 1.87~eV to 2.38~eV,
respectively. In Figure~\ref{fig:tuning}b) both, QDs and barrier, emission
energies are plotted against one another.

For a quantum box, the lower potential of the confining barrier, the lower energies of the confined levels. For the case of QDs, the smaller band gap of a barrier and/or a QD material, the smaller energy of optical transitions of the QD confined excitons. Expected decrease of the QD emission energy with the decreasing barrier energy is indeed evidenced in Fig.~\ref{fig:tuning}b). Despite of this clear trend, a scatter of QDs emission energies is present in Fig.~\ref{fig:tuning}b). It reflects differences in the QD layer growth parameters (e. g., an amount of CdSe deposited or
deposition temperature) in the case of a different samples. Thus, the
Fig.~\ref{fig:tuning}b) confirms that decreasing of the (Zn,Cd)Se barrier
energy by the increase of Cd content provides an efficient way of tuning of
the CdSe QDs energy.
The lowest (Zn,Cd)Se achieved barrier energy in the sample set is equal to
2.46~eV, what determines~\cite{Mourad:PRB2010} the largest obtained x$_{Cd}$ in
the barrier to 30\% $\pm$ 3\%, as intended on the sample growth stage.
A further increasing of the Cd content in the barrier would result in a further lowering of the QD energies. We estimate that x$_{Cd}$ = 80\% would still ensure at least single QD confined level.

The Fig.~\ref{fig:tuning}b) clearly shows that, in particular, decreased
barrier energy enables CdSe QDs with energy of emission lower then the
Mn$^{2+}$ ion internal transition energy. The resulting hindering of the
non-radiative exciton recombination related to Mn$^{2+}$ ions is highly
advantageous for studies and implementation of Mn-doped CdSe QDs
\cite{Worschech:SSciTech2008, Beaulac:NanoLett2008,
Chernenko:PRB2005, Lee:PRB2005, Chekhovich:PRB2007}. Moreover, the QDs
emitting in yellow spectral range (2.05~eV - 2.15~eV) are promising for
implementation in lasers operating in the short wavelength transmission
window of plastic optical fibers \cite{Murofushi:POF1996}.

\section{Quasi-resonantly excited individual CdSe/(Zn,Cd)Se Quantum Dots}
In most of the studied samples, in the low energy tail of the QD emission
band there are spectrally resolved emission lines attributed to individual
QDs. Resonant or quasi-resonant excitation enables further limiting of the
number of excited QDs and, in consequence, optical accessing of individual
QDs. It opens a perspective for coherent manipulation of QD confined
excitonic states \cite{Bonadeo:Science1998, Flissikowski:PRL2001}, or
studies of spin memory effects on electron or ion confined to the QD
\cite{Goryca:PRL2009, Suffczynski:PRB2008, Smolenski:PRB2015}.

\begin{figure}[!t]
\includegraphics[width=0.9\linewidth]{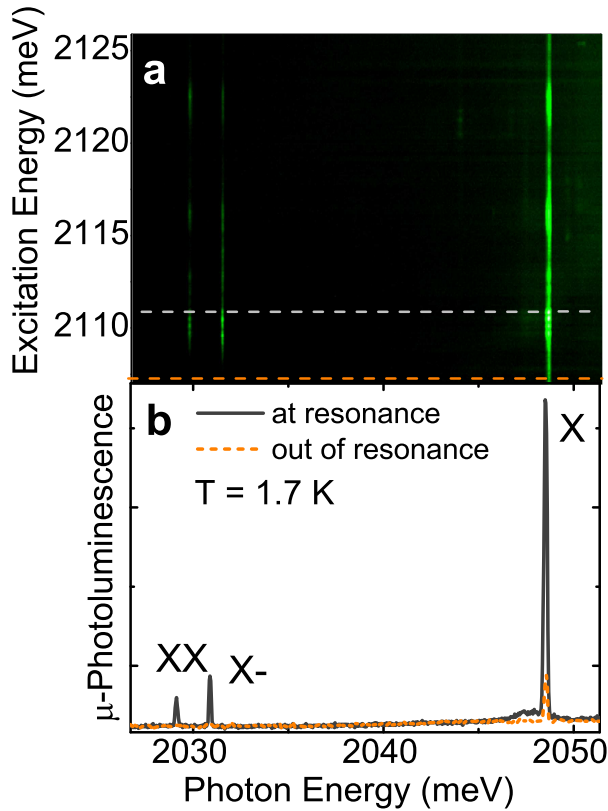}
\caption{(a) Photoluminescence Excitation spectra of a selected
CdSe/(Zn,Cd)Se QD (QD1) at 2105-2127~meV excitation energy range. (b)
$\mu$-PL spectra of the QD1 at cross-sections of the panel a), as indicated
by dashed lines for resonant (E$_{exc}$ = 2110.8~meV) and non-resonant
(E$_{exc}$ = 2107.5~meV) excitation case.}
\label{fig:PLE}
\end{figure}

Figure~\ref{fig:PLE} presents the result of the PLE measurement on a
selected CdSe/(Zn,Cd)Se QD (QD1). A map composed of emission spectra of the
QD1 acquired for excitation energies increasing consecutively with a step
of 72~$\mu$eV in a range from 2107~meV to 2127~meV is shown in
Figure~\ref{fig:PLE}a).
A set of sharp (FWHM down to 0.15~meV) maxima is evidenced in the PLE map
in Fig.~\ref{fig:PLE}a) for each excitonic transition of the QD1. The
emission spectra of QD1 excited quasi-resonantly (at 2110.8 meV) and out of
resonance (at 2107.5 meV) are displayed in Figure~\ref{fig:PLE}b). Much
larger intensity of the emission for the quasi-resonant excitation confirms
increased efficiency of transfer of excitation to the QD as compared to
non-resonant case.
a quasi-resonant transfer of excitation to the emitting QD1 state.

\begin{figure}
\includegraphics[width=1.1\linewidth]{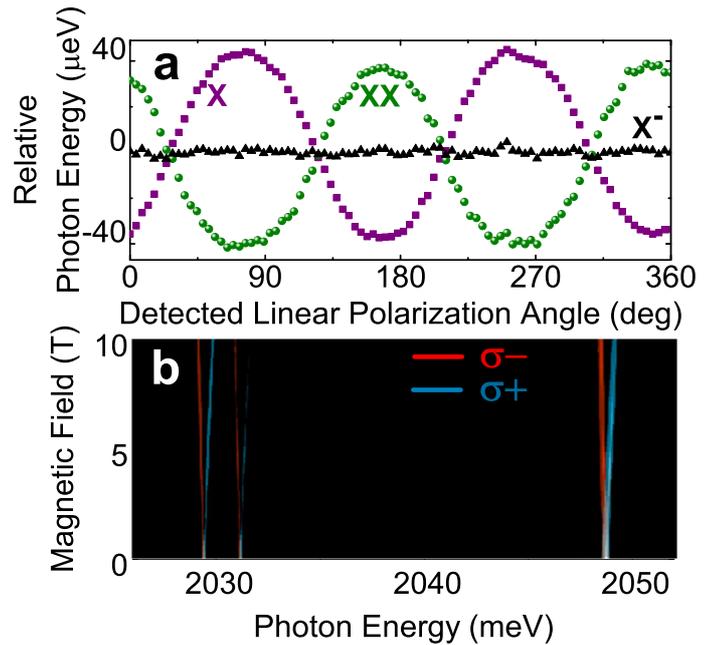}
\caption{a) Relative emission energy of the X, X$^-$ and XX confined in the
QD1 as a function of a detected linear polarization angle. b) Emission
spectrum of the QD1 in the magnetic field of up to 10~T applied in Faraday
configuration. Transitions in $\sigma+$ and $\sigma-$ circular
polarizations are encoded by blue and red color, respectively.}
\label{fig:characterization}
\end{figure}

The QD1 transitions are identified by measurements of anisotropy of linear
polarization of emission (see Fig.~\ref{fig:characterization}a)) as a
recombination of neutral exciton (X), biexciton (XX) and charged exciton
(X$^-$). Exciton anisotropy splitting is determined to 80~$\mu$eV. The
negative sign of the charged exciton is determined in PLE experiment, as
described below. Energy difference between X and XX transitions is equal to
19.4~meV, the value that is common for the QDs in the studied set of
samples and generally quite typical for CdSe QDs \cite{Kulakovskii:PRL1999,
Smolenski:PRB2015}. The variation of transition energy of the QD1 with
magnetic field applied in Faraday configuration (see
Fig.~\ref{fig:characterization}b)) yields excitonic g-factor equal to 1.2,
slightly smaller than for typical CdSe/ZnSe QDs \cite{Kulakovskii:PRL1999,
Puls:PRB1999}, and diamagnetic shift $\gamma$ = 1.2~$\mu$eV/T$^2$
comparable to the one observed previously \cite{Hundt:pssb2001}.

As it is seen in Figure~\ref{fig:PLE}a), no difference in resonance energy
for X$^-$ and neutral excitons is found, in contrary to the case of previously
studied (In,Ga)As QDs \cite{Finley:PRB2001}. Instead, the resonance
energies are common for the X, X$^-$ and XX transitions of QD1. A similar
conduct has been observed previously in the case of CdTe/ZnTe QDs\cite{Kazimierczuk:PRB2009, Goryca:PRL2009}, where it has been interpreted
as a result of coupling between two neighboring QDs: the one at higher
energy absorbing the excitation, and the emitting one at lower energy, to
which the excitation is transferred. In the case of the presently
investigated QDs, however, the absorbing state does not exhibit optical
properties expected for an absorbing QD observed previously
\cite{Kazimierczuk:PRB2009}, like anisotropy exchange splitting with two
orthogonal polarization axes and linear dependence of energy on the
magnetic field due to the Zeeman effect (not shown). Energy distance
between PLE maxima and QD1 transitions does not match to any multiple of
LO-phonon energy in CdSe (26.5~meV), what excludes also the possibility of
excitation via LO-phonon \cite{Hundt:pssb2001, Flissikowski:PRL2001,
Scheibner:PRB2003}. Exact assessment of the nature of the absorbing states
in the present case requires further studies.

\begin{figure}
\includegraphics[width=0.9\linewidth]{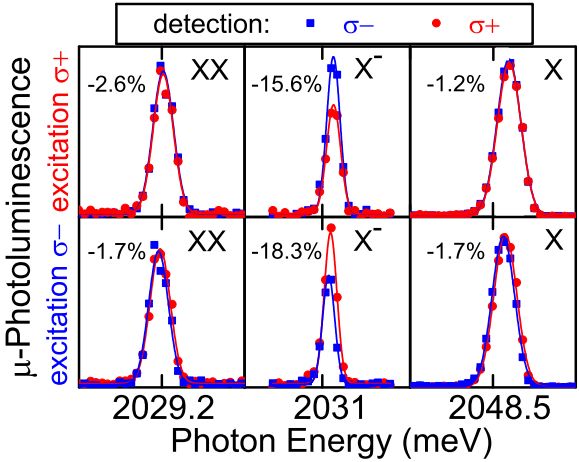}
\caption{The emission spectra of resonantly excited X, X$^-$ and XX
transitions of QD1 for circularly polarized excitation and detection along
with a degree of polarization transfer (see text), as indicated. The width
of the energy window equals to 1~meV in each of the plots.}
\label{fig:transfer}
\end{figure}

In order to investigate a transfer of exciton spin polarization between the
absorbing and the emitting state, we perform polarization resolved PLE
measurements. We record four PLE maps corresponding to four combinations of
circular polarizations of the excitation and detection of the light.
Resonantly excited emission spectra of X, X$^-$ and XX transitions of the
QD1 obtained in that way are presented in Fig.~\ref{fig:transfer}.

In order to quantify the amount of exciton spin polarization conserved in
the transfer process, we introduce degree of the polarization transfer $P$.
The $P$ is defined for polarization of the excitation $\sigma+$ as:

\begin{equation}
P_{\sigma+}=\frac{I_{\sigma+/\sigma+} - I_{\sigma+/\sigma-}}
{I_{\sigma+/\sigma+} + I_{\sigma+/\sigma-}},
\label{Eqn:poltransfer}
\end{equation}

where $I_{\sigma+/\sigma-}$ denotes the intensity of excitonic transition
excited in $\sigma+$ and detected in $\sigma-$ polarization. In analogy,
the $P_{\sigma-}$ is defined as $P_{\sigma-} = (I_{\sigma-/\sigma-} -
I_{\sigma-/\sigma+})/ $ $ (I_{\sigma-/\sigma-} + I_{\sigma-/\sigma+})$.
The values of $P_{\sigma+}$ and $P_{\sigma-}$ obtained for X, X$^-$ and XX
transitions of QD1 are indicated in respective panels in
Fig.~\ref{fig:transfer}. As it is seen, the X and XX transitions exhibit a
negligible transfer of polarization. It is expected, since exchange
interaction in the anisotropic QD mixes spin-up and spin-down exciton
states, which leads to a linearly polarized eigenstates
\cite{Bayer:PRB2002}. As a result, the initial polarization of the neutral
exciton is erased after it is confined to the QD.

The Fig.~\ref{fig:transfer} indicates, however, that in the case of the
X$^-$ transition the cross-polarized emission is favored with respect to
the co-polarized one. The $P_{\sigma+}$ and $P_{\sigma-}$ attains the value
of -15.6\% and -18.3\%, respectively. The maximum value of $P$ obtained for
the X$^-$ in the studied QDs is -26.1\%. As it has been described in
detail, e.g., in Refs.~\cite{Cortez:PRL2002, Ware:PRL2005}, the mechanism
of such efficient, negative polarization transfer \cite{Cortez:PRL2002,
Ware:PRL2005, Murayama:PRB2007, Kazimierczuk:PRB2009, Smolenski:PRB2015}
results from electron-heavy hole spin flip-flop process in triplet state of
charged exciton, being the intermediate state of the transfer to the
emitting singlet state. Within such interpretation, the heavy-hole spin
relaxation rate is much faster than the one of electron, and the electron
is a majority carrier in the charged exciton (X$^-$) complex.
QD a

\begin{figure}
\includegraphics[width=0.9\linewidth]{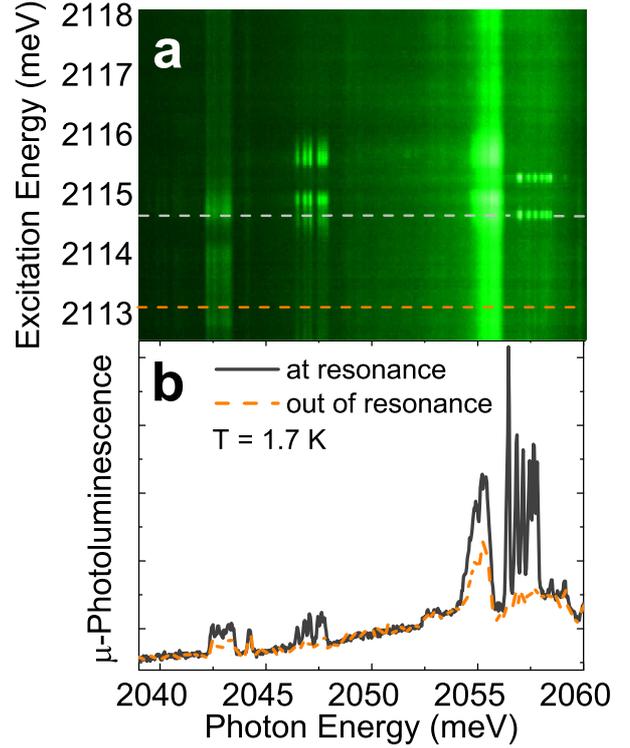}
\caption{(a) PLE spectra of individual CdSe/(Zn,Cd)Se QD embedding a single
Mn$^{2+}$ ion. (b) $\mu$-PL spectra of QD2 under resonant (E$_{exc}$ =
2114.7 meV) and non-resonant (E$_{exc}$ = 2113.1 meV) excitation.}
\label{fig:Mn}
\end{figure}

The QD embedding the single Mn$^{2+}$ ion exhibits a characteristic
six-fold splitting of the exciton emission line \cite{Kobak:NatureCom2014,
Besombes:PRL2004, Goryca:PRL2009}. An example PLE spectrum along with
respective $\mu$-PL spectra providing the evidence for a quasi-resonant
optical addressing of QDs that contain the single Mn$^{2+}$ ion are shown
in Fig.~\ref{fig:Mn}. Similar emission patterns in the case of the line sets observed at energies of 2043~meV, 2047~meV, 2057~meV suggest that they are related to excitonic complexes (e.g., X, X$^{-}$ and XX) coming from the same QD. However, for QDs without the Mn$^{2+}$ ion in the studied samples, resonance energies are common for all excitonic complexes confined in the same QD (see the example case of the QD1 in Fig.~\ref{fig:PLE}). Thus, the Fig.~\ref{fig:Mn} with different resonance energy for each of the four transitions, would suggest that four QDs embedding the Mn$^{2+}$ ion contribute to the PLE and $\mu$-PL spectra.

Presence of four QDs with the single Mn$^{2+}$ ion under the 0.5~$\mu$m excitation spot would indicate the average planar density of such QDs equal to 50/$\mu$m$^2$. Such value is at least two orders of magnitude higher than the estimated one. In any case, the locally increased density of QDs embedding individual Mn ions has been observed for
several regions within the sample. The inhomogeneity of spatial
distribution of such QDs suggests a positive correlation between Mn ions
spatial positions within the QD layer arising at the sample growth stage,
in particular during self-assembled QD formation. It may result from a kind
of aggregation of the Mn$^{2+}$ ions prior to their adsorption on the
sample surface during the MBE growth. More detailed studies are required in
order to confirm and to determine the driving mechanism here.

\section{Conclusions}
Fabrication and properties of resonantly excited emission of CdSe QDs
embedded in (Zn,Cd)Se barrier are presented. The (Zn,Cd)Se barrier enables
lowering of energy of QD confined levels (down to 1.87~eV) with respect to
a typical CdSe/ZnSe QDs (2.32~eV). This should facilitate studies and
applications of selenium based QDs doped with several Mn$^{2+}$ ions and
for their implementation as active material in lasers operating in the
short wavelength transmission window of plastic optical fibers. Observed efficient transfer of polarization in the studied QDs is valuable
for the field of quantum information processing, where stored spins may
play a role of quantum memories or qubits controlled and measured optically
\cite{Imamoglu:PRL1999, Cortez:PRL2002, Suffczynski:PRB2008}.

\section*{Acknowledgments}
This work was supported by the Polish National Research Center (NCN)
projects under numbers \\
UMO-2013/10/E/ST3/00215, DEC-2013/09/B/ST3/02603,
DEC-2011/02/A/ST3/00131, and MNiSW project Iuventus Plus IP2014 034573.

\section*{References}

\bibliography{CdSe_QDs_biblio}

\end{document}